\documentclass{amsart}

\usepackage{amsmath, amsthm, amssymb, graphics, graphicx}
\usepackage{subfigure}
\usepackage{hyperref}
\graphicspath{%
    {converted_graphics/}
    {Images/}
}
\newcommand{\mathsym}[1]{{}}
\newtheorem{theorem}{Theorem} 
\newtheorem{lemma}{Lemma}
\newtheorem{corollary}{Corollary}
\newtheorem{definition}{Definition}
\newtheorem*{conjec}{Conjecture}
\newcommand{\PSSPM}{PSSPM}
\newcommand{\SPM}{SPM}
\newcommand{\SSPM}{SSPM}
\newcommand{\PSPM}{PSPM}
\newcommand{\Ater}{Alt}
\newcommand{\pater}{PAlt}
\newcommand{\atom}{Atom}

\usepackage{graphicx}
\begin{document}
\title[Fixed-point forms of the parallel symmetric sandpile model]{Fixed-point forms of the parallel symmetric\\ sandpile model}
\author{E. Formenti}
\address{University of Nice-Sophia Antipolis, Department of Informatics, Parc Valrose, 06108 Nice Cedex 2, France}
\email{Enrico.FORMENTI@unice.fr}
\author{V.T. Pham}
\address{Institute of Mathematics, 18 Hoang Quoc Viet, Hanoi, Vietnam}
\email{pvtrung@math.ac.vn}
\author{H.D. Phan}
\address{Institute of Mathematics, 18 Hoang Quoc Viet, Hanoi, Vietnam and Liafa, University of Denis Diderot-Paris 7, Case 7014-2, Place Jussieu 75256 Paris, Cedex 05, France}
\email{phanhaduong@math.ac.vn and phan@liafa.jussieu.fr}
\author{T.T.H. Tran}
\address{Institute of Mathematics, 18 Hoang Quoc Viet, Hanoi, Vietnam}
\email{ttthuong@math.ac.vn}




\begin{abstract}
This paper presents a generalization of the sandpile model, called the parallel symmetric sandpile model, which inherits the rule of the symmetric sandpile model and implements them in parallel. We prove that although the parallel model produces really less number of fixed points than that by the sequential model, the forms of fixed points of the two models are the same. Moreover, our proof is a constructive one, which gives a nearly shortest way to reach a given fixed point form. 
\end{abstract}
\date{}
\maketitle
\section{Introduction}\label{S:intro}
Sandpile model ($\SPM$) was introduced by Bak, Tang and Weisenfeld \cite{BTW87} as a paradigm to describe the self-organized criticality (SOC) phenomenon in physics and has a variety of applications in physics, mathematics,  economics, theoretical computer science \cite{BTW87, Bak99, Big99, BLS91, DRSV95, GMP02b, GMP02c, LMMP01, LP00, RC00}.  The simplest $\SPM$ model is that the system starts from a single column configuration $(n)$, then at each step, one column gives one grain to its right neighbor if it has more than at least two grains comparing to its right neighbor. It was proved \cite{GLMMP01, GMP02b, LMMP01} that this model converges to only one configuration at which the evolution rule can not be applied at any column (this configuration is called \emph{fixed point}). Furthermore, all \emph{reachable configurations} (which are obtained from the initial configuration $(n)$ by applying several times of the evolution rule) are also well characterized and its configuration space is a lattice.  The system $\SPM$ has been modified and generalized in several aspects to satisfy each particular purpose. In the context of chip firing games, cellular automata and informatics systems, the $\SPM$ model with parallel update scheme (\emph{i.e.} at each step, all applicable rules are applied in parallel) received great attention \cite{BG92, CFM07,  Dur98}. In \cite{Dur98}, Durand-Lose showed that the transient time to reach a fixed point is linear in the total number of grains $n$ when the parallel updated scheme is used, whereas it is $\mathcal O(n^{3/2})$ when the sequential one is used.

To make it closer to the real physical phenomenon, Formenti \emph{et al.}~\cite{FMP07} and Phan~\cite{Pha08}, generalized $\SPM$ so that grains are allowed to fall on both sides (left and right). This generalized model  is called \emph{symmetric sandpile model} and denoted by $\SSPM$. The model has no unique fixed point any more. While Formenti \emph{et al.} investigated the model by considering its configurations without caring its positions (that is, they identify all configurations which are up-to a translation on a line), Phan investigated the model in addition to its positions and showed the furthest position (comparing to the position at which the initial column is situated). The authors characterized reachable configurations in ~\cite{FMP07} (resp. forms of reachable configurations in ~\cite{Pha08}) starting from a single column configuration. Furthermore, they showed that the number of fixed-point forms of the model is exactly $[\sqrt n]$. 

In this paper, we study the $\SSPM$ model using a parallel update scheme. We denote this by $\PSSPM$ for further ease of reference. We stress that unlike $\SPM$ and $\PSPM$, $\SSPM$ and $\PSSPM$ can have multiple fixed points when starting from a given initial configuration (as usual, we only consider single column configurations). It is also remarkable that in $\PSPM$ it is difficult to characterize all reachable configurations and so it is the same result in $\PSSPM$. In this paper, we concentrate on fixed points of the parallel scheme $\PSSPM$. It is straightforward that the set of fixed points of  $\PSSPM$ is contained in the set of fixed points of $\SSPM$. We show as main result in this paper that although this containment is proper in general, the containment of their forms of fixed points is an equality. Therefore, we can obtain all forms of fixed points of the sequential model by using the parallel update scheme with less than time. The proof is long and involved, and it has been divided into several subparts for better understanding. Indeed, the paper is structured as follows. Section 2 represents definitions and results about the characterizations of reachable configurations as well as the time of convergence for the three models: $\SPM$, $\PSPM$ and $\SSPM$. Section 3 contains the main result. First, we give a precise definition of $\PSSPM$. Then, we introduce three procedures which are the building blocks for the proof of the main result (Theorem \ref{T:PSSPM1}). Finally, in Section 4 we draw the conclusion and present some perspectives for future research on the subject.

\section{Sandpile model and some generalizations}\label{S:spm}
In this section, we first give some basic definitions related to integer partitions and unimodal sequences. Then we represent the results of the sandpile model ($\SPM$) and two generalizations ($\PSPM$ and $\SSPM$) investigated in \cite{Dur98, FMP07, GLMMP01, GMP02b, Pha08}.
\begin{definition} Let $n$ and $k$ be positive integers. Then
\begin{itemize}
\item[(i)] An \emph{integer partition} is a non-increasing sequence of positive integers $a=(a_1,a_2,\dots,a_k)$, moreover if $a_1+\cdots+a_k=n$ then $a$ is called a \emph{partition} of $n$.
\item[(ii)] A \emph{unimodal sequence} of length $k$ is a sequence of $k$ positive integers $(a_1,a_2,\dots, a_k)$ such that there exists an index $1\leq i\leq k$ satisfying the condition $a_1\leq a_2\leq \cdots \leq a_i \geq a_{i+1}\geq \cdots\geq a_{k-1}\geq a_k$. The quantities defined by $$h(a)=\max \{a_i\}_{i=1}^k \ \ \ \text{ and } \ \ w(a)=\sum_{i=1}^k a_i$$ are respectively called \emph{the height} and \emph{the weight} of $a$.
\item[(iii)] \emph{The reserve of a sequence} $a=(a_1,a_2,\dots, a_k)$, denoted by $a^{-1}$, is the sequence $(a_k,a_{k-1},\dots, a_1)$.
\item[iv)] A \emph{$n$th power} of a sequence of positive integer $a$, denoted by $a^n$, is the sequence obtained by concatenating $n$ times $a$.  
\end{itemize}
\end{definition}
Given a unimodal sequence $a$ and an index $1\leq i\leq k$, we denote  $$a_{<i}=(a_1,\dots,a_{i-1}) \text{ and } a_{>i}=(a_{i+1},\dots,a_k),$$
$$a_{\leq i}=(a_1,\dots,a_{i-1},a_i) \text{ and } a_{\geq i}=(a_i,a_{i+1},\dots,a_k),$$ and call them the \emph{strict left sequence}and the \emph{strict right sequence} of $a$ by $i$, the \emph{left sequence} and the \emph{right sequence} of $a$ by $i$, respectively. 

A discrete dynamical system is described by its \emph{configurations} and its \emph{evolution rule}. A configuration $b$ is \emph{reachable} from another configuration $a$ if $b$ is obtained from $a$ by applying several times the evolution rule, and we write $a\to b$. We usually consider the system starting from one configuration, called the \emph{initial configuration}, and then we investigate the set of all configurations reachable from this initial one, and we call this set the \emph{configuration space} of the system. By this way, the system is well defined by its evolution rule and its initial configuration. A \emph{fixed point} (or \emph{stable configuration}) of the model is a configuration reachable from the initial one and on which the evolution rule can not be applied.

The \emph{sandpile system} is a discrete dynamical system used to describe the self-organized criticality (SOC) phenomena in physics. Our studying models are in a subclass of the sandpile system. In these models, a configuration, also called \emph{sandpile}, is represented by an integer sequence $(a_1,a_2,\dots,a_k)$, the part $a_i$ is called the \emph{height} of the pile (or column) $i$. In this paper, we always assume that for all sandpile models, the initial configuration is a single column containing $n$ grains and the position at which these grains is situated is called its \emph{initial column}. Therefore, each model is well-defined by its evolution rule. We first give here the definition of the simplest sandpile model ($SPM$) introduced in \cite{BLS91}.

\begin{definition}[\cite{BLS91}] The \emph{Sand pile model} is a system defined by the following \emph{$\SPM$ rule} (right rule):
\begin{itemize}
\item column $i$ is \emph{right collapsible} if $a_i-a_{i+1}\geq 2$ and when it collapses on the right it gives one grain to its right neighbor;  
\item at each step, there is at most one collapse.
\end{itemize}
\end{definition}
\begin{figure}[h]
\centering
\includegraphics[width=6cm]{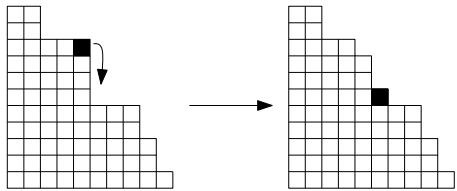}
\caption{$SPM$ rule}
\end{figure}
\begin{figure}[h]
\includegraphics[height=6cm]{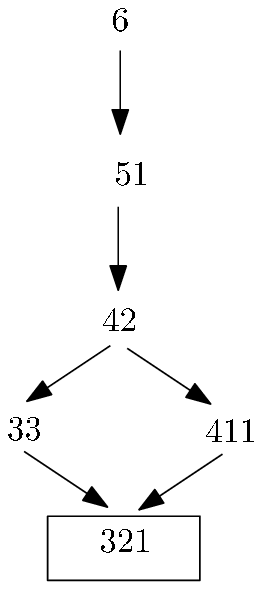}\hfill 
\includegraphics[height=6cm, width=6cm]{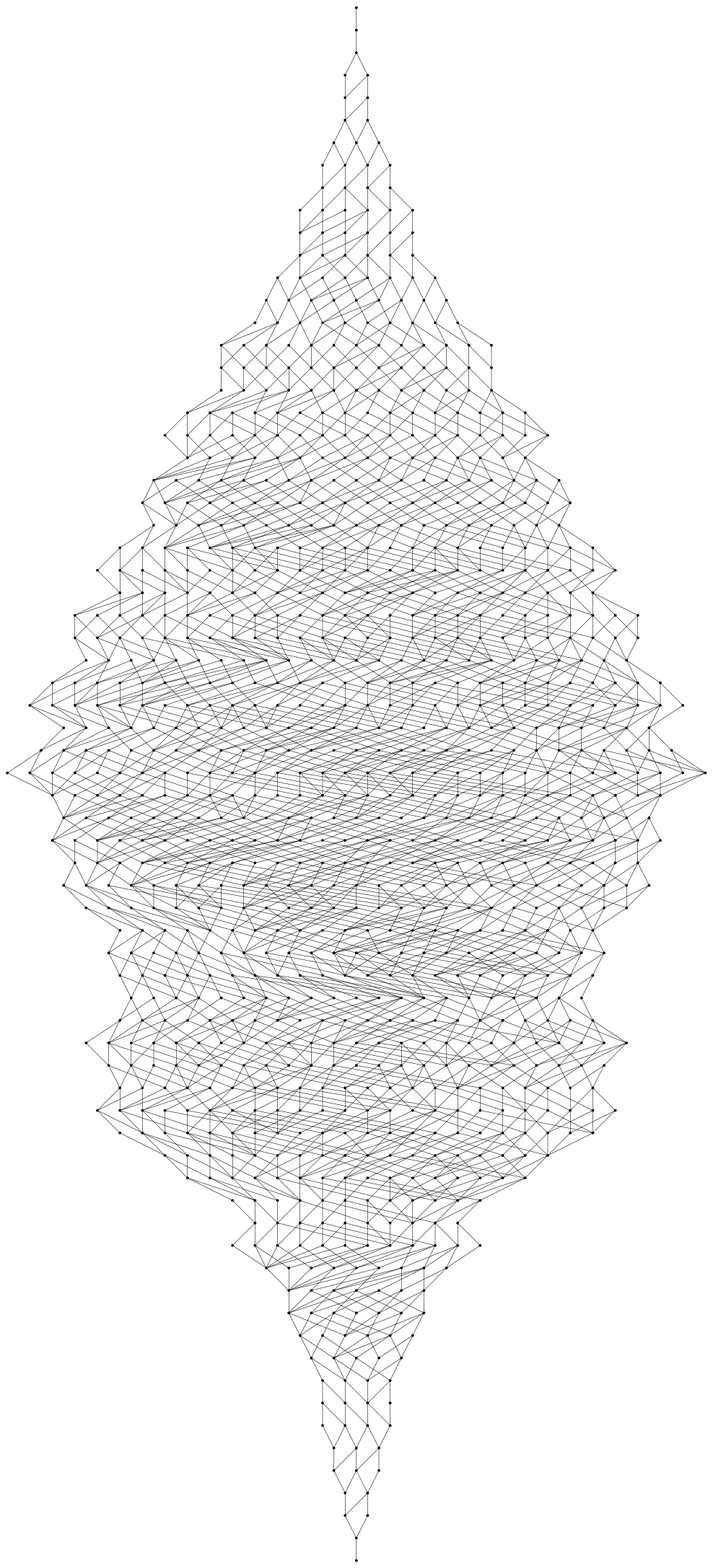}
\caption{Configuration spaces $SPM(6)$ and $SPM(30)$}
\end{figure}
We denote by $\SPM(n)$ (resp. $\SPM$) the configuration space of the sandpile model starting from $(n)$ (resp. from any single column configuration). Reachable configurations of $\SPM$ are characterized as the following:

\begin{theorem}[\cite{GLMMP01}]\label{T:GMP1} Let $c$ be an integer partition. Then $c$ is a configuration of $\SPM$ if only if it does not contain any
 subsequence of the form $(p,p,p)$ or $(p,p,p-1,p-2,....,q+1,q,q)$ for positive integers $p,q$  satisfying $0<q<p$.
\end{theorem}
From this characterization, the fixed point of $\SPM$ is given by an explicit formula.
\begin{corollary}[\cite{GLMMP01}]\label{C:GMP2} Given a positive integer $n$. Then $\SPM(n)$ has a unique fixed point which is of the form $(p,p-1,p-2,....,q,q,q-1,....2,1)$ with $p,q\in \mathbb N$ and $q\leq p$. Furthermore, the time to reach this fixed point is $\mathcal O (n^{3/2})$.
\end{corollary}

Now, we define the $\SPM$ in parallel (so called \emph{parallel sandpile model}) introduced by Durand-Lose \cite{Dur98}.  
\begin{definition}[\cite{Dur98}] The parallel sandpile model is a system defined by the following \emph{$PSPM$ rule}: 
\begin{itemize}
\item at each step, each column collapses at most once;
\item at each step, all columns which are right collapsible collapse on the right.
\end{itemize}
\end{definition}

We also denote by $\PSPM(n)$ (resp. $\PSPM$) the configuration space of the parallel sandpile model starting from the single column configuration $(n)$ (resp. any single column configuration). 

\noindent{\bf Remark:}
\begin{itemize}
\item[-] In $SPM$ and $PSPM$, the first column is always a highest column. 
\item[-] The $SPM$ is non-deterministic (since at each step, there may exist several collapsible columns) whereas the $PSPM$ is deterministic. Both models have the same unique fixed point. 
\item[-] Although it is impossible to reach all reachable configurations of the sequential $SPM$ by parallelism, it is possible to get its fixed point in linear time by parallelism.
\end{itemize}
\begin{figure}[h]
\centering
\subfigure[Configuration space $\SPM(6)$]
{\label{spm6}\includegraphics[width=2.5cm]{spm6}}
\hspace{1.5cm}
\subfigure [Configuration space $\PSPM(6)$]
{\hspace{3cm}\label{pspm6}
\includegraphics[height=4.5cm]{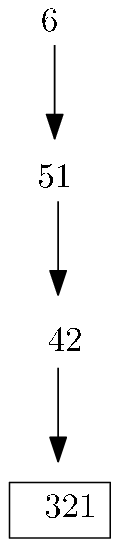}%
}
\caption{}
\end{figure}
The following result gives us the time to reach the fixed point by parallel model. It is $\mathcal O(n)$ comparing to $\mathcal O(n^{3/2})$ in sequential model.
\begin{theorem}[\cite{Dur98}] \label{T:pspm} Let $n$ be a positive integer. The time to reach the fixed point from $(n)$ in $\PSPM(n)$ is $\mathcal O(n)$.
\end{theorem}

Next, we represent a recent generalization of $\SPM$ where columns not only collapse on the right but also on the left. 
\begin{definition}[\cite{FMP07, Pha08}] \emph{The symmetric sandpile model} is a system defined by \emph{SSPM rule} as follows:
\begin{itemize}
\item addition to the right rule in $\SPM$ model, there is also the left rule, that mean one column $i$ can give one grain to its left neighbor if $a_i-a_{i-1}\geq 2$. 
\item at each step, there is at most one collapse.
\end{itemize}
\end{definition}
We denote by $\SSPM(n)$ (resp. $\SSPM$) the configuration space of the symmetric sandpile model starting from $(n)$ (resp. any single column configuration).  

\noindent{\bf Remark:}
\begin{itemize}
\item[-] The $\SSPM$ is a non-deterministic model since there may have columns which are collapsible on both sides and it may have more than one fixed point.
\item[-] Unlike $\SPM$, each configuration of $\SSPM$ is a unimodal sequence.
\item[-] Unlike $\SPM$ and $\PSPM$ where the initial column is always a highest one, the position of the highest column of $\SSPM$ can be changed during the evolution. 
\end{itemize}
\begin{figure}[h]
\centering
\includegraphics[width=8cm]{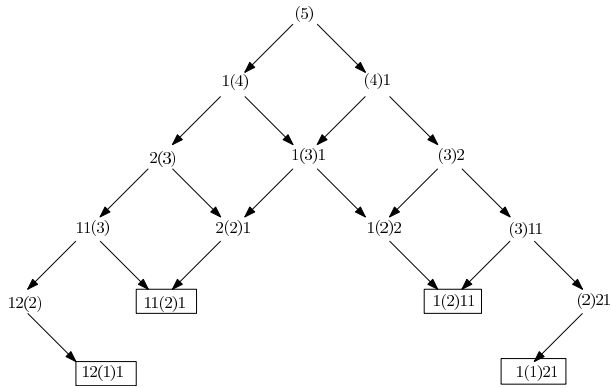}
\caption{Configuration space $\SSPM(5)$}
\label{sspm6}
\end{figure}

Figure \ref{sspm6} illustrates all the reachable configurations of $\SSPM$ starting with $5$ grains, where the initial column is  in the parentheses. Then, $\SSPM(5)$ has $4$ fixed points as shown in the figure. It is also important to consider the fixed-point forms which does not mention to their positions. Hence, we identify all fixed points which are up-to a translation. For instance, $11(2)1$ and $1(1)21$ have the same form $1121$ (but different in the position); $12(1)1$  and $1(2)11$ have the same form $1211$. Hence, fixed points of $\SSPM(5)$ are in one of two forms $1121$ and $1211$. The study of form of $\SSPM$ is investigated in \cite{FMP07} and that of position is in \cite{Pha08}. In fact, these papers give a characterization of the form of $\SSPM$.
\begin{theorem}[\cite{FMP07, Pha08}]\label{T:FMP1}
A unimodal sequence $a=(a_1,a_2,...,a_{k-1},a_k)$ is the form of a reachable configuration of $\SSPM$ if only if there exists an index $1\leq i\leq k$ such that the sequences $a_{\geq i}$ and $(a_{<i})^{-1}$ are configurations of $\SPM$.
\end{theorem}
Furthermore, the enumeration of fixed-point forms of $\SSPM$ is also given.
\begin{theorem}[\cite{FMP07}] \label{T:FMP2} The number of fixed-point forms of $SSPM(n)$ is $[\sqrt{n}]$. Moreover, if $P$ is a fixed point of $\SSPM(n)$ then $P$ is of height either $[\sqrt n]$ or $[\sqrt n]-1$.
\end{theorem}
\section{The parallel symmetric sand pile model}\label{S:psspm}
In this section we introduce another generalization of the sandpile model. In this model, we inherit the rule of the symmetric sandpile model and implement them in parallel. 

First we give precisely its definition. Like the other generalizations of the sandpile model represented in the previous section, we always start with the single-column configuration. 
\begin{definition} \emph{The parallel symmetric sandpile model} is a system defined by the following \emph{$\PSSPM$ rule}: 
\begin{itemize}
\item At each step, all collapsible columns collapse;
\item For columns which are collapsible on both sides, it must choose exactly one direction to collapse.
\end{itemize}
\end{definition}

\begin{figure}[h]
\centering
\includegraphics[width=5cm]{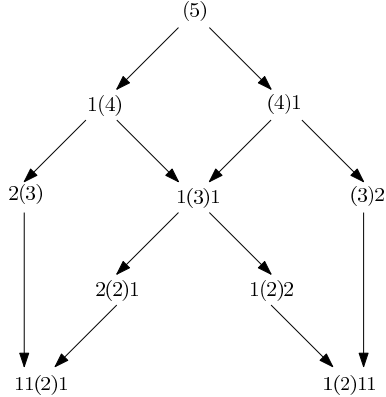}
\caption{Configuration space $\PSSPM(5)$}
\label{psspm6}
\end{figure}

We denote by $\PSSPM(n)$ (resp. $PSSPM$) the configuration space of the parallel symmetric sandpile model starting with $(n)$ (resp. any single column configuration).

\noindent{\bf Remark:} 
\begin{enumerate}
\item Unlike $\PSPM$ which is deterministic, the $\PSSPM$ is non-deterministic since although columns collapse at the same time at each step, there may have two directions (must choose one) for one column collapsing;
\item Since an evolution step by $\PSSPM$ rule can be considered as a combination of some evolution steps by $\SSPM$ rule, each configuration of $\PSSPM$ is a $\SSPM$ configuration. Furthermore, the configuration space of $\PSSPM$ is a subspace of that of $\SSPM$ and the set of fixed points of $\PSSPM(n)$ is a subset of that of $\SSPM(n)$.
\end{enumerate}

We notice that the set of fixed points of $\PSSPM$ is a proper subset of that of $\SSPM$. Actually, $\SSPM(5)$ has $4$ fixed points: $12(1)1$, $11(2)1$, $1(2)11$, $1(1)21$ (see Figure \ref{sspm6}), but $\PSSPM(5)$ has only $2$ fixed points: $1(2)11$ and $11(2)1$ (see Figure \ref{psspm6}). However, one can observe that $\SSPM(5)$ has only $2$ fixed-point forms as $\PSSPM(5)$, which raises a question about the correlation of fixed-point forms of the two models. 

The main result of this paper is to state that the set of fixed-point forms of $\PSSPM$ and that of $\SSPM$ are the same. Moreover, we can show an explicit evolution by $\PSSPM$ rule to reach any given fixed-point form of $\SSPM$. 
 

\begin{theorem}\label{T:PSSPM1}
The set of fixed-point forms of $\PSSPM(n)$ is equal to that of $\SSPM(n)$. Consequently, there is $[\sqrt{n}]$ fixed-point forms of $\PSSPM(n)$.
\end{theorem}
As our proof lengthens in many steps, we break the proof into some lemmas. First, we sketch the main idea of the proof, we leave its details at the end of this paper after presenting necessary procedures. 

\begin{proof}[Main idea]
For a fixed point $P$ of $\SSPM(n)$, we construct a sequence of $PSSPM$ transitions to obtain $P$ from the initial configuration $(n)$. Because we are interested in the form of $P$ but not in its position, we can suppose that the center column of $P$ is at position $0$ (the notion of ``center column", one of highest columns, will be given later). In the constructed evolution, the column $0$ is always a highest one, so the choice of $PSSPM$ rule in each step is in fact the choice of direction collapsing of the column $0$.  

For a symmetric fixed point $P$, \emph{i.e.} $\left(P_{<0}\right)^{-1}=P_{>0}$. The evolution is an Alternating Procedure, described as follows: at odd steps, the column $0$ collapses on the right, and at even steps, it collapses on the left. From $(n)$ this procedure will converge to the symmetric fixed point $P$ (see Corollary \ref{C:sym}). 

For $P$ not symmetric, we can suppose that the column $0$ is the center of $P$, \emph{i.e.} 
$$d=|w(P_{>0})-w(P_{<0})|=\underset{i}{\min}|w(P_{>i})-w(P_{<i})|.$$ Not loosing generality we assume that $w(P_{>0})-w(P_{<0})>0$ and let $d=w(P_{>0})-w(P_{<0})$. The evolution by $\PSSPM$ rule is composed of three procedures:
\begin{itemize}
\item[i)] Pseudo-Alternating Procedure: a procedure from $(n)$ to the configuration $Q=\big(1,2,\dots, d-1, (n-d^2), d,d-1,\dots, 2,1\big)$  (see Lemma \ref{L:pse1}). Note that $w(Q_{>0})-w(Q_{<0})$ is exactly $d$.
\item[ii)] Alternating Procedure: a procedure from $Q$ to the configuration $R$ on which we could not apply any more the Alternating Procedure. It will be proved that $R$ is of height $h$ (see Lemma \ref{L:alp1}).
\item[iii)] Deterministic procedure: a deterministic procedure from $R$ to $P$, where at each its step there is no any column collapsible on both sides (see Lemma \ref{L:fin}). 
\end{itemize}
Like the symmetric case, we claim that at the end of the evolution we obtain $P$ by Lemma \ref{L:pse3}, Lemma \ref{L:alp1} and Lemma \ref{L:fin}. 
\end{proof}
To present the proof we first introduce some necessary procedures. The following procedure is implemented on integer partitions.
\begin{definition}[Atom Procedure]
Let $t$ be a non-negative integer and let $a$ be an integer partition. The \emph{Atom Procedure of $t$ steps} is a sequence of $t$ transitions starting from $a$ described as follows:
\begin{itemize}
  \item[(i)] The $\PSPM$ rule is applied at all steps.
  \item[(ii)] At each odd step one grain is added to the first column and at each even step no grain is added.
\end{itemize}
\end{definition}
\begin{figure}[h]
\includegraphics[width=12cm, height=4.5cm]{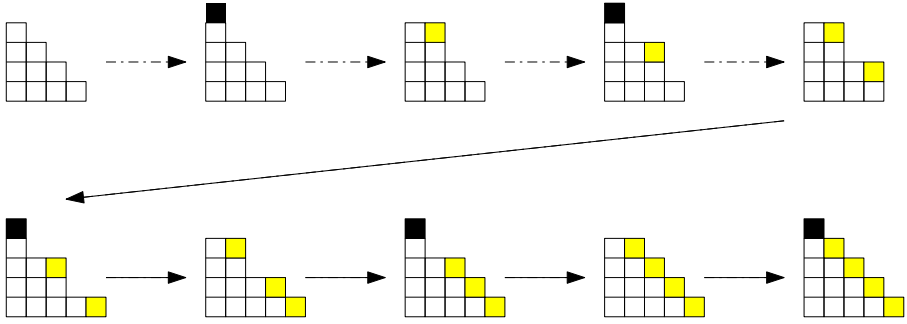}
\caption{9 first steps of Atom Procedure from $(4,3,2,1)$. The dark grains are added at odd steps}
\label{atom4}
\end{figure}
We denote by $\atom^t(a)$ the configuration obtained from $a$ after $t$ steps of Atom Procedure.

Recall that each sequential sandpile model is coded by a chip firing game on line by taking differences between two consecutive parts of each integer partition. By this we associate each integer partition $a=(a_1,\dots,a_k)$ to the sequence of its differences $d(a)=(d_1,\dots,d_k)$, where $d_i=a_{i}-a_{i+1}$ (by convention $a_{k+1}=0$). Therefore if $a_{i}-a_{i+1}\geq 2$ then $d_i\geq 2$. Moreover, if $a^\prime$ is obtained from $a$ by one transition by $\SPM$ rule at position $i$ then $d(a^\prime)=(d_1,\dots,d_{i-1}+1,d_i-2,d_{i+1}+1,\dots,d_k)$.

Denote by $s(k)=(k,k-1,\dots,2,1)$ the stair of $k$ steps. The following lemma describes all configurations which are reachable from $s(k)$ by Atom Procedure.
\begin{lemma}\label{L:Atom1}
Let $k,t$ be positive integers such that $0\leq t\leq 2k+1$. The following statements hold
\begin{itemize}
\item[(i)] If $0\leq t\leq k$ then $d\left(\atom^t(s(k))\right)$ is of the form $((0,2)^{\frac {t}{2}},1^{k-t})$ if $t$ is even and of the form $(2,(0,2)^{\frac{t-1}{2}}, 1^{k-t})$ if $t$ is odd.
\item[(ii)] If $k+1\leq t\leq 2k+1$ then $d\left(\atom^t(s(k))\right)$ is of the form $(0,(2,0)^{k+1-\frac{t-1}{2}},1^{t-k})$ if $t$ is odd and of the form $((2,0)^{k+1-\frac t2},1^{t-k})$ if $t$ is even.
\end{itemize}
Consequently, the height of $\atom^t(s(k))$ is equal to $k+1$ if $t$ is odd and $k$ if $t$ is even.
\end{lemma}
\begin{proof}[Proof]
We prove this lemma by induction on $t$. For $t=0$ we have $\atom^0(s(k))=(k,k-1,\dots,2,1)$ so that $d\left(\atom^0(s(k))\right)=(1^{k})$ which corresponds to the statement.

For $t=1$ then by definition of Atom Procedure one sand grain is added to the first column. The new configuration is $\atom^1(s(k))=(k+1,k-1,\dots,2,1)$ and $d\left(\atom^1(s(k))\right)=(2,1^{k-1})$ which corresponds to the statement. We assume that the statement holds till step $t$. We prove that it also holds for the step $t+1$. We consider the following cases:

\emph{Case 1}: If $t\ne k$ and $t$ is odd. We have $$d\left(\atom^t(s(k))\right)=(2,(0,2)^{\frac {t-1}{2}}, 1^{k-t})  \text{ if } t < k$$ and  $$d\left(\atom^t(s(k))\right)=((2,0)^{k+1-\frac t2},1^{t-k})  \text{ if }  t\geq k+1.$$ Since $t+1$ is even, we does not add any grain at this step and just apply the $\PSPM$ rule. Hence, $$d\left(\atom^{t+1}(s(k))\right)=((0,2)^{\frac{t+1}{2}},1^{k-t-1})\text{ if }  t<k$$ and $$d\left(\atom^{t+1}(s(k))\right)=(0,(2,0)^{k+1-\frac {t+1}{2}},1^{t-k+1}) \text{ if } t\geq k+1,$$ respectively.

\emph{Case 2}: If $t\ne k$ and $t$ is even. By the definition of Atom Procedure one grain is added on the first column at the step $t+1$. It transforms $(0,2)^{\frac {t}{2}},1^{k-t}$ (resp. $(2,(0,2)^{\frac{t-1}{2}}, 1^{k-t})$) into $(0,(2,0)^{k+1-\frac{t-1}{2}},1^{t-k})$ (resp. $((2,0)^{k+1-\frac t2},1^{t-k})$).

\emph{Case 3}: If $t=k$. We have then $$d\left(\atom^{t}(s(k))\right)=((0,2)^{k}) \text{ if }  t  \text{ is odd }$$ and $$d\left(\atom^{t}(s(k))\right)=(0,(2,0)^{\frac{k+1}{2}})  \text{ if } t \text{ is even.}$$ We also have $$\left(\atom^{k+1}(s(k))\right)=(2,(0,2)^{\frac{k}{2}},1)  \text{ and } $$ $$ \left(\atom^{k+1}(s(k))\right)=((2,0)^{\frac{k}{2}},1), \ \ \text{ respectively}.$$ So the forms of the associated configurations are described as the statement.

Now it is remarkable that if the associated sequence of a configuration has the form $((0,2)^l,1^m)$ then the configuration is of height $2l+m$. So the rest of the lemma is straightforward.
\end{proof}
\begin{corollary}\label{C:Atom2} The Atom Procedure transforms the configuration $(k,k-1,...,2,1)$ into the configuration $(k+1,k,...,2,1)$ after $2k+1$ transitions.
\end{corollary}

Next, we introduce two other procedures which are implemented on unimodal sequences
\begin{definition}[Alternating Procedure] Let $a$ be a unimodal sequence whose highest column is placed at position $0$ and this column has a large enough number of grains to enable to distribute to its neighbors. An \emph{Alternating Procedure} is a sequence of transitions by $\PSSPM$ rule such that
\begin{itemize}
  \item[(i)] Column $0$ collapses on the right at odd steps.
  \item[(ii)] Column $0$ collapses on the left at even steps.
\end{itemize}
\end{definition}

\noindent{\bf Remark:}
\begin{enumerate}
\item For a given configuration, the Alternating Procedure will be not implemented forever. It stops when the column $0$ has not enough grains to collapse either on the left at some even step or on the right at some odd step.

\item At all steps in Alternating Procedure, all columns except column $0$ have only one direction to collapse due to the unimodality. Furthermore, at each step, the column $0$ choose exactly one direction to collapse. Therefore, the configuration obtained from $a$ after $t$ steps of Alternating Procedure is uniquely determined.

\item The column $0$ decreases exactly $1$ after one step of Alternating Procedure. Consequently, after $t$ steps of this procedure the column $0$ decreases exactly $t$ grains.

\end{enumerate}
\begin{figure}[h]
\centering
\includegraphics[width=7cm, height=5cm]{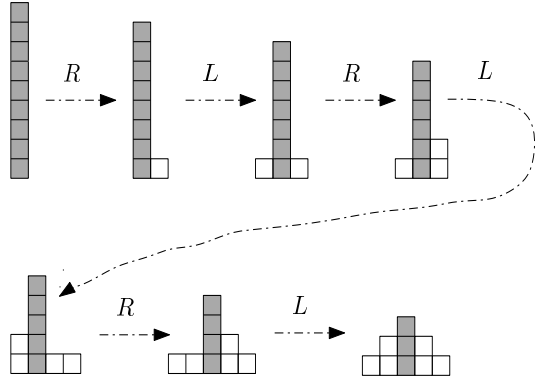}
\caption{6 first steps of Alternating Procedure from (9). The arrow together with the direction $R$ or $L$ (Right or Left) corresponding to the direction along which the column 0 (dark column) collapses}
\label{alt9}
\end{figure}
Figure \ref{alt9} illustrates that we can not implement more than $6$ steps of Alternating Procedure from (9). 

Denote by $\Ater^t(a)$ the configuration obtained from $a$ after $t$ steps of Alternating Procedure. The following facts are straightforward
$$\left(\Ater^{t}(a)\right)_{>0}=\atom^{t}\left(a_{>0}\right),$$ and 
$$\big(\big(\Ater^{t}(a)\big)_{<0}\big)^{-1}=\atom^{t-1}\big(\big(a^\prime_{<0}\big)^{-1}\big),$$
where $\big(a^\prime_{<0}\big)^{-1}$ is obtained from $\big(a_{<0}\big)^{-1}$ by applying one step by $\PSPM$ rule. In other word, applying $t$ steps of the Alternating Procedure on $a$ is the same as applying $t-1$ steps of Atom Procedure on $a^{\prime}_{<0}$ and $t$ steps of Atom Procedure on $a_{>0}$.  

Particularly, if we start from the singe-column configuration then after some steps of Alternating Procedure the weights of the strict left part and strict right part by $0$ differ at most 1. 

We have the following lemma

\begin{lemma}\label{L:alt1} Let $n,k$ be positive integers such that $n\geq k^2$ and let $a=(1,2,...,k-1,n-k^2+k,k-1,...,2,1)$ be a unimodal sequence. Then $a$ is reachable from $(n)$ after $k^2-k$ steps of Alternating Procedure.
\end{lemma}
\begin{proof}[Proof]
We prove this statement by induction on $k$. For $k=1$ then the configuration $(n)$ is reachable from itself. We assume that the configuration $(1,2,\dots,t-1,n-t^2+t,t-1,\dots,2,1)$ is reachable from $(n)$ after $t^2-t$ steps of Alternating Procedure. That means $\Ater^{t^2-t}(n)=(1,2,\dots,t-1,n-t^2+t,t-1,\dots,2,1)$. We need to show that the unimodal sequence $(1,2,\dots,t,n-t^2-t,t,\dots,2,1)$ is reachable from $(n)$ by Alternating Procedure (where $n\geq (t+1)^2$). 


Since $$\big(\Ater^{t^2-t}((n))\big)_{<0}=\big(\Ater^{t^2-t}((n))\big)_{>0}=s(t-1),$$ by Corollary \ref{C:Atom2} and the above remark it is sufficient to show that the Alternating Procedure can be implemented on $\Ater^{t^2-t}(a)$ in $2t$ steps.

Thus, for $1\leq i\leq 2t$ then by Lemma \ref{L:Atom1} we have 
$$h\big(\atom^{i-1}\big(\Ater^{t^2-t}(a)\big)_{<0}\big)\leq t,$$ and 
$$h\big(\atom^{i-1}\big(\Ater^{t^2-t}(a)\big)_{>0}\big)\leq t$$
In addition, due to $n\geq (t+1)^2$ we have
$$n-(t^2-t)-(i-1)\geq n-(t^2-t)-(2t-1)=n-t^2-t+1\geq t+2.$$
So after $i-1$ steps of Alternating Procedure the column $0$ of $\Ater^{t^2-t}(a)$ is still collapsible. This completes the proof.

\end{proof}
\begin{corollary}\label{C:sym} Let $k$ be positive integer. Then the unimodal sequence $(1,2,\dots,k-1,(k),k-1,\dots,2,1)$ is reachable from $(k^2)$ in $\PSSPM$ by Alternating Procedure.
\end{corollary}
Let $s(n,k)=(1,2,\dots,k-2,k-1,(n-k^2),k,k-1,\dots,2,1)$ be a unimodal sequence whose the column $0$ is of height $n-k^2$, the left strict part by $0$ is a stair of $k-1$ steps and the right strict part by $0$ is a stair of $k$ steps. We have the following lemma 

\begin{lemma} \label{L:ALT2} Let $n,k$ be positive integers such that $n\geq (k+1)^2+(k+1)$. Then the Alternating Procedure transforms the configuration $s(n,k)$ into the configuration $s(n,k+1)$ after $2k+1$ steps.
\end{lemma}
\begin{proof} By Lemma \ref{L:Atom1}, it is sufficient to prove that the column $0$ of $s(n,k)$ has a large enough number of grains to enable to collapse in $2k+1$ steps. We prove this by induction on the number of implemented steps. For $1\leq t\leq 2k$ we have
$$h\left(\atom^t(s(n,k)_{>0})\right)\leq k+1 \ \ \text{ and } \ \ h\big(\atom^t\big((s(n,k)_{<0})^{-1}\big)\big)\leq k+1.$$
On the other hand, since $n\geq (k+1)^2+(k+1)$ we have 
\begin{align*}
n-k^2-(t-1)&\geq n-k^2-(2k-1)\\
& \geq (k+1)+2\\
&\geq h\big(\atom^t(s(n,k)_{>0})\big)+2,
\end{align*} and
\begin{align*}
n-k^2-(t-1)&\geq (k+1)+2\\
&\geq h\left(\atom^t\left(\left(s(n,k)_{<0}\right)^{-1}\right)\right)+2.
\end{align*}
It implies that the column $0$ of $s(n,k)$ can collapse at the step $t$. That means the Alternating Procedure can implement in $2k$ steps.

Moreover, by Lemma \ref{L:Atom1} we have $$\left(\Ater^{2k}(s(n,k))\right)_{>0}=\atom^{2k}(s(n,k)_{>0})=(k,k,\dots,2,1)$$ and $$\left(\left(\Ater^{2k}(s(n,k))\right)_{<0}\right)^{-1}=\atom^{2k-1}\left(\left(s(n,k)_{<0}\right)^{-1}\right)=(k,k-1,\dots,2,1),$$ (since $\left(s(n,k)_{<0}\right)^{-1}$ is stable, after one step by $\PSPM$ rule on $(s(n,k)_{<0})^{-1}$ we obtain itself).

Besides, $$h\left(\Ater^{2k}(s(n,k))\right)=(n-k^2)-2k\geq k+2=h\left(\left(\Ater^{2k}(s(n,k))\right)_{>0}\right)+2.$$
So that we can apply the Alternating Procedure on $s(n,k)$ in $2k+1$ steps and obtain $s(n,k+1)$. 
\end{proof}
Now we define the second procedure implemented on unimodal sequences.
\begin{definition}[Pseudo-Alternating Procedure] Let $t$ be a positive integer and let $a$ be a unimodal sequence. We assume that the highest column of $a$ is placed at position $0$ and this column has a large enough number of grains to enable to distribute to its neighbors. The \emph{Pseudo-Alternating Procedure} of $t$ steps on $a$ is a sequence of $t$ transitions by $\PSSPM$ rule starting from $a$ such that Alternating Procedure is applied from step $i^2+1$ to step $(i+1)^2$ (for  $i=0,1,\dots,[\sqrt t]-1$) and from step $[\sqrt t]^2+1$ to step $t$.
\end{definition}
\begin{figure}[h]
\includegraphics[width=11cm, height=6cm]{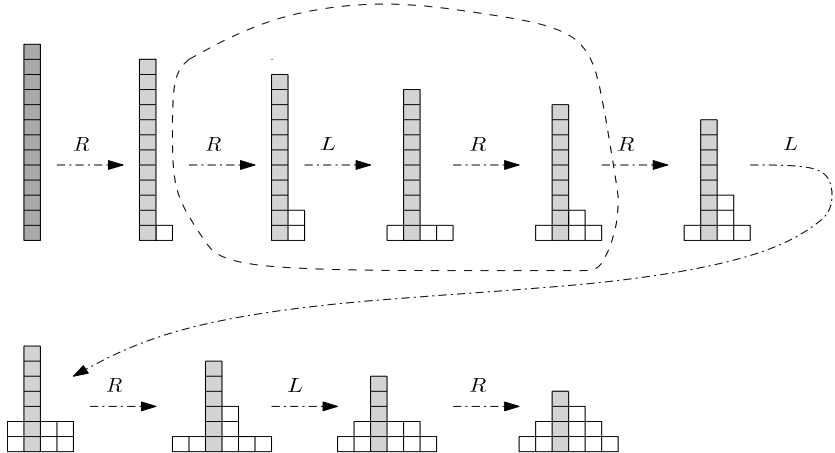}
\caption{9 first steps of Pseudo-Alternating Procedure on (13)}
\label{pseu13}
\end{figure}
Remark that a Pseudo-Alternating Procedure is a concatenation of Alternating Procedure, however it is not an Alternating Procedure (see Figure \ref{pseu13}).

Denote by $\pater^{t}(a)$ the configuration obtained from $a$ after $t$ steps of Pseudo-Alternating Procedure.

Figure \ref{pseu13} shows $9$ steps of Pseudo-Alternating Procedure starting from $(13)$. The dashed area illustrates that Alternating Procedure is applied from the step $2$ to step $4$ corresponding to the case $i=1$ in the above definition. It is also noticeable that we are not able to continue implementing Pseudo-Alternating Procedure on $(13)$ more than $9$ steps although after 9 first steps the darked column of height $4$ is still collapsible on the left but it is un-collapsible on the right. 

\begin{lemma}\label{L:pse1} Let $n,k$ be positive integers such that $n\geq (k+1)^2+(k+1)$. The Pseudo-Alternating Procedure transforms the configuration $(n)$ into the configuration $(1,2,...,k,n-(k+1)^2,k+1,k,...,2,1)$ after $(k+1)^2$ steps.
\end{lemma}
\begin{proof}[Proof]
We prove this by induction on $k$. For $k=1$ then the configuration $(n-1,1)$ is reachable from $(n)$, where the column $0$ collapses on the right by Pseudo-Alternating Procedure. We assume that the configuration $(1,2,\dots,t-1,n-t^2,t,t-1,\dots,2,1)$ is reachable from $(n)$ after $t^2$ steps of Pseudo-Alternating Procedure ($t\leq k$). We can write $\pater^{t^2}=(1,2,\dots,t-1,n-t^2,t,t-1,\dots,2,1)$. We need to prove that the configuration $a=(1,2,\dots,t,n-(t+1)^2,t+1,\dots,2,1)$ is also reachable from $(n)$ if  $n\geq (t+1)^2+(t+1)$. Due to the determination of the Pseudo-Alternating Procedure it is equivalent to prove that $a$ is obtained from $\pater^{t^2}((n))$ after $2t+1$ steps of Alternating Procedure. This is actually obtained from Lemma \ref{L:ALT2}. 
\end{proof}
Now given a fixed-point form  $P=(p_1,p_2,\dots,p_k)$ of $\SSPM$, we put $$Div^i(P)=|w(P_{<i})-w(P_{>i})|,$$ and 
$$Div(P)=\min_{1\leq i\leq k}\{Div^i(P)\}.$$ We call $Div(P)$ the \emph{symmetric difference} of $P$. A column $i$ of $P$ at which $Div^i(P)$ gets minimum is called the \emph{symmetric separator} of $P$.
\begin{figure}[h]%
\includegraphics[width=10cm]{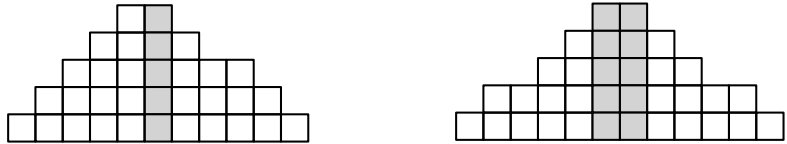}%
\caption{Two fixed-point forms $P=(1,2,3,4,5,5,4,3,3,2,1)$ and $P^\prime=(1,2,2,3,4,5,5,4,3,2,2,1)$ with their separators marked by  dark columns, and $Div(P)=2$; $Div(P^\prime)=5$}%
\label{sepa}%
\end{figure}

\noindent{\bf Remark:} A fixed-point form may have more than one symmetric separator. For instance, the form $(1,2,2,3,4,5,5,4,3,2,2,1)$ contains two symmetric separators: column 6 and 7 (see Figure \ref{sepa}).

We have the following lemma
\begin{lemma} \label{L:div} Let $P=(p_1,p_2,\dots,p_k)$ be a fixed-point form of $\SSPM$ and $h=h(P)$. Then there exists an index $i$ of $P$ such that $p_i=h$ and $Div^i(P)\leq h$. Moreover, for this position $i$ we have $Div^i(P)=Div(P)$.
\end{lemma}
\begin{proof} Put $t=\min\{i: p_i=h\}$. Since $P$ is a fixed point of $\SSPM$, we have
$$\frac{h(h-1)}{2}\leq w(P_{<t})\leq \frac{h(h-1)}{2}+h-1,$$ and
$$\frac{h(h-1)}{2}\leq w(P_{>t})\leq \frac{h(h-1)}{2}+3h.$$
So that $0\leq |P_{<t}-P_{>t}|\leq 3h$. We consider the following cases

{\it Case 1.} If $0\leq |P_{<t}-P_{>t}|\leq h$ then $t$ is the position satisfying the statement.

{\it Case 2.} If $h+1\leq |P_{>t}-P_{<t}|\leq 3h$, we claim that $t+1$ is the position satisfying the statement. First we show that $p_{t+1}=h$. On the contrary $p_{t+1}\leq h-1$, we have $$\frac{h(h-1)}{2}\leq w(P_{>t})\leq \frac{h(h-1)}{2}+h-1.$$ Therefore, $|P_{>t}-P_{<t}|\leq h-1$ which contradicts the condition $|P_{<t}-P_{>t}|\geq h+1$. So that we must have $p_{t+1}=h$ and this also implies that $$P_{>t}\geq \frac{h(h-1)}{2}+h\geq P_{<t}.$$ On the other hand, we have $P_{<(t+1)}=P_{<t}+h$ and $P_{>(t+1)}=P_{>t}-h$. Hence $$|P_{>(t+1)}-P_{<(t+1)}|=|P_{>t}-P_{<t}-2h|\leq h.$$ 
Last, we prove that $Div^i(P)=Div(P)$ by showing that for other positions $t$ then $Div^t(P)\geq h$. Thus, it is sufficient to show this is true for positions $t$ satisfying that $p_t\leq h-1$. Since the strict part of $P$ (which contains the column of height $h$) by $t$  is of weight at least $\frac{h(h+1)}{2}$ corresponding to the sequence $(h,h-1,\dots,1)$ and the rest strict part by $t$ is of weight at most $\frac{h(h-1)}{2}$ corresponding to the sequence $(h-1,\dots,2,1)$. We have $$Div^t(P)\geq \frac{h(h+1)}{2}-\frac{h(h-1)}{2}\geq h.$$ This completes the proof.
\end{proof}

\begin{figure}[h]
\includegraphics[width=12cm, height=4cm]{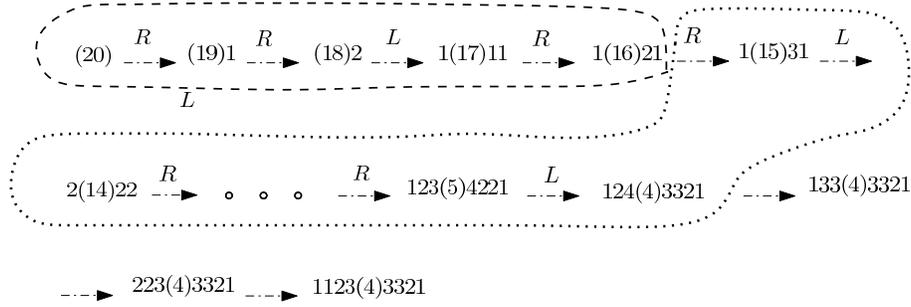}
\caption{A way of length $19$ to obtain the fixed point $(1,1,2,3,(4),3,3,2,1)$ from $(20)$}
\label{wayfixed}
\end{figure}

Now we present the proof of Theorem \ref{T:PSSPM1}
\begin{proof}[Proof in details of Theorem \ref{T:PSSPM1}]
We recall here some notations used in the main idea of the proof Theorem \ref{T:PSSPM1}: $$d=Div(P)\ \ \text{ and }\ \ h=h(P).$$ Now we represent precisely the construction the sequence of transitions by $\PSSPM$ rule to obtain $P$ from $(n)$ placed at $0$.
\begin{enumerate}
  \item[(1)] Applying the Pseudo-Alternating Procedure from the step $0$ to the step $d^2$.
  \item[(2)] Applying the Alternating Procedure from the step $d^2+1$ to the step $n-h$.
  \item[(3)] Applying the deterministic procedure from the step $n-h+1$ (from this step there is no any column collapsible on both sides).
\end{enumerate}
It is noticeable that the construction showed above also works for both $P$ symmetric and not symmetric. We recall that we are not always able to apply the Pseudo-Alternating Procedure and the Alternating Procedure. The next is to prove that we can implement $d^2$ steps of Pseudo-Alternating Procedure on $(n)$; then we can implement $(n-h-d^2)$ steps of Alternating Procedure on $\pater^{d^2} (n)$; last, we get $P$ at the end of the evolution. These statements will be proved in the following three lemmas.
\begin{lemma} \label{L:pse3}It is possible to implement $d^2$ steps of Pseudo-Alternating on $(n)$. 
\end{lemma}
\begin{proof} By Lemma \ref{L:pse1}, it is sufficient to show that $n\geq d^2+d$. Since $P$ is a fixed point of height $h$ and of symmetric difference $d$, so $P$ must contain at least $h^2+d$ grains corresponding the configuration $(1,2,\dots,h-1,h,h-1,\dots,d+1,d,d,\dots,2,1)$. By Lemma \ref{L:div}, we have $d\leq h$ and so $n\geq d^2+d$ and $$\pater^{d^2}\left((n)\right)=s(n,d).$$
\end{proof}
\begin{lemma} \label{L:alp1}It is possible to implement $(n-h-d^2)$ steps of Alternating Procedure on $\pater^{d^2}\left((n)\right)$. 
\end{lemma}
\begin{proof} 
On the contrary we assume that $t$ is the first step at which this does not hold and $1\leq t\leq n-d^2-h$. By Lemma \ref{L:Atom1} and the definition of Alternating Procedure, we have the following facts \begin{align*}
\left(\Ater^i(s(n,k))\right)_{0}&=n-d^2-i,\\
\big(\Ater^i(s(n,k))\big)_{-1}\leq h+1  &\ \text{ and } \    \big(\Ater^i(s(n,k))\big)_{1}\leq h+1, 
\end{align*}
for all 
$i=1,2,\dots,t-1$.

Put $\Ater^{t-1}(s(n,d))=b$. Since we can not implement the Alternating Procedure on $\pater^{d^2}((n))$ at the step $t$, we must have
$$b_0 \leq b_{-1}+1 \leq h+2 \ \ \text{ in case } t \text{ is even}$$ and
$$b_0 \leq b_{1}+1 \leq h+2 \ \ \text { in case } t \text{ is odd}.$$  
Combining with the hypothesis of the contrary $t\leq n-d^2-h$, we deduce for both cases that $$h+1\leq n-d^2-t+1=b_0\leq h+2.$$ Hence, either $n-d^2-t+1=h+1$ or $n-d^2-t+1=h+2$.

On the other hand, we have $w(P_{>0})-w(P_{<0})=d$ and $w(P_{<0})+w(P_{>0})+h=n$, so that $n-d-h$ is even. Thereby, $n-d^2-t+1=h+1$ if $t$ is even and $n-d^2-t+1=h+2$ if $t$ is odd. We consider these two cases

{\it Case 1.} $t$ is even. We have $b_0=h+1$  and $b_{-1}\geq h.$ Since there is no grain added on the left of $\Ater^{t-2}\big(s(n,d))$ at the step $t-1$, we get  $$\big(\Ater^{t-2}\big(s(n,d))\big)_{-1}\geq h.$$
It is remarkable that in the Alternating Procedure, the strict left and the strict right by $0$ sequentially fulfill the stairs before creating the new stairs of greater length. So that as from the step $d^2+t-2$ the strict left part is at least the stair $(1,2,\dots,h)$. This is also true for the strict right part by $0$ since its weight is more than that of the strict left part. So during the evolutions of applying the transitions arbitrarily by the $\PSSPM$ rule on $b$, the column $0$ of $b$ never collapses again. Therefore, at the end of these evolutions, we get fixed points having the same height $h+1$. We will show this is a contradiction. We assume that $P^\prime$ is a such fixed point. Then $w(P^{\prime}_{>0})-w(P^\prime_{<0})=d+1,$ and by Theorem \ref{T:FMP1}, $P$ and $P^\prime$ are of representation as follows
$$P^\prime=\big(1,2,\dots,\alpha^\prime,\alpha^{\prime}\dots,h+1,\dots,\alpha^\prime+d+1,\alpha^\prime+d+1,\dots,2,1\big)$$
and
$$P=\big(1,2,\dots,\alpha,\alpha\dots,h,\dots,\alpha+d,\alpha+d,\dots,2,1\big).$$ 
Since $w(P^{\prime})=w(P)$ and $P^\prime_0=P_0+1$, it implies that $\alpha^\prime=\alpha-1$ and $P^\prime_{-1}=P^{\prime}_1=h$. So that $P_1=P_{-1}=P_0=h$ and $P$ contains at least 3 different plateaus of 4 following plateaus: $P_{-h+\alpha-2}P_{-h+\alpha-1}$ of height $\alpha$, $P_{-1}P_0$ of height $h$, $P_0P_1$ of height $h$ and $P_{h-\alpha-d+1}P_{h-\alpha-d+2}$ of height $\alpha+d$ (and $h-\alpha-d\geq 0$). This contradicts the condition that $P$ is the form of a fixed point of $SSPM$.

{\it Case 2.} $t$ is odd. We have $$b_0=h+2\ \text{ and }\ b_{1}=h+1.$$ Since there is no grain added on the right of $\Ater^{t-2}\big(s(n,d)\big)$ at step $t-1$, we have $$\big(\Ater^{t-2}(s(n,d))\big)_{1}\geq h+1.$$ 
Therefore,
$$w(\Ater^{t-2}(s(n,d))_{j>0})\geq \frac{(h+1)(h+2)}{2},$$ and 
\begin{align*}
w(P)=w(\Ater^{t-2}(s(n,d)))&\geq (h+1)(h+2)-d+h+2\\
&\geq h^2+4h+4-d > h^2+3h.
\end{align*} This contradicts the condition that $P$ is one fixed point of $\SSPM(n)$ and of height $h$ (since $h^2+3h$ corresponds to the configuration  $(1,2,\dots,h-1,h,h,h,h,h-1,\dots,2,1)$).
\end{proof}
In summary, we conclude that starting from $(n)$ we can implement the Pseudo-Alternating Procedure in $d^2$ steps then the Alternating Procedure in $n-d^2-h$ steps. As a result we obtain the unimodal sequence whose all columns are of height less than or equal to $h$.

\begin{lemma} \label{L:fin}The procedure applying the transitions by $\PSSPM$ rule on the configuration obtained after the processes $(1)$ and $(2)$ above is deterministic. Moreover, it converges to $P$. 
\end{lemma}
\begin{proof}
This is straightforward from the fact that all columns of the configuration obtained after two above processes are of the height less than $h$. Furthermore, the column $0$ of the configurations in the procedure is always of the height $h$. Therefore, there is no any columns collapsible on both sides and the procedure is deterministic. On the other hand, the left side part and the right side part by $0$ evolve independently by $PSPM$ (so it is deterministic) to reach their unique fixed points. The difference of two these parts always is $d$. So at the end of the procedure we get $P$. 
\end{proof}
The three above lemmas have ended our proof of the theorem.
\end{proof}
Figure \ref{wayfixed} shows one way to obtain the fixed point $(1,1,2,3,(4),3,3,2,1)$ (of form $112343321$) from the single column configuration of $20$ grains. The dashed zone illustrates $4$ steps of Pseudo-Alternating Procedure to obtain the $s(20,2)$ (which is $1(16)21$). The dotted zone illustrates $12$ next steps of Alternating Procedure to obtain the configuration of height $4$ at column $0$ (equal to the height of the given fixed point). Last, it needs to apply $3$ steps of $\PSSPM$ rule when there is no column collapsible on both sides to obtain the final fixed point.  

We also remarkable that the evaluations of inequalities in Lemmas \ref{L:alt1}, \ref{L:ALT2},\ref{L:pse1} and specially Lemma \ref{L:alp1} are very sensible. Furthermore, our procedures we constructed above are not commutative each other in general, especially when the center column has not enough grains to distribute to its neighbors. So this may be easy to lead to another fixed-point form of $\SSPM$ not the one we expect. For instance, for the form $P=(122221)$ we have $h(P)=2, \ d(P)=2$ and if we start from $(10)$ and implement the way shown in proof of Theorem \ref{T:PSSPM1} then we get exactly $P$. But if we first do the Alternating Procedure in $4$ steps (which is equal to $n-d^2-h$) and next do the Pseudo-Alternating Procedure, then we are able to implement this procedure only in $3$ steps not $4$ steps (which is equal to $d^2$). Hence, we final get the form $123211$ which is not $122221$. 

Last we give a upper bounded for the shortest length of the ways to reach a fixed point in $\PSSPM$. 
\begin{corollary} Let $T_{\PSSPM}(n)$ be the shortest length to reach a fixed point in $\PSSPM(n)$. Then $n-[\sqrt n]\leq T_{\PSSPM}(n)\leq n.$
\end{corollary}
\begin{proof}
Recall that if $h$ is the height of a fixed point of $\PSSPM(n)$ then $h=[\sqrt n]$ or $[\sqrt n]-1$. Hence, $T_{\PSSPM(n)}\geq n-h$.  From the way we constructed above, it takes $n-h$ transitions of applying Pseudo-Alternating Procedure and Alternating Procedure; then it takes at most $h$ transitions of applying the final procedure to reach a fixed point of $\PSSPM(n)$. Therefore, $T_{\PSSPM(n)}\leq n$. 
\end{proof}
\nocite{Sta98}
\section{Conclusion and Perspective}\label{S:conclusion}

We proved that beginning with a singleton column of sand grains, the sequential model and the parallel model produce the same  fixed-point forms. To tackle the problem, for each fixed-point form of $\SSPM$, we construct an explicit way of $\PSSPM$ evolution to obtain this fixed-point form. Every configuration in this way has a ``smooth'' form, even it can be characterized by a formula on the time of the evolution, whereas it is difficult to capture the forms of general reachable configurations. 

Actually, the problem of finding a shortest way to reach a given fixed-point form of $\SSPM$ is interesting to explore. The way we constructed is not always a shortest way although it reveals many interesting properties to be possibly a shortest way. In fact, the difference between the length of our constructed way and the one of the shortest ways is at most $[\sqrt n]$. We do not know so far an explicit formula of the length or the behavior of such shortest ways. 
\begin{figure}[h] 
  \centering
  \includegraphics[bb=25 79 341 235,width=3.59in,height=1.77in,keepaspectratio]{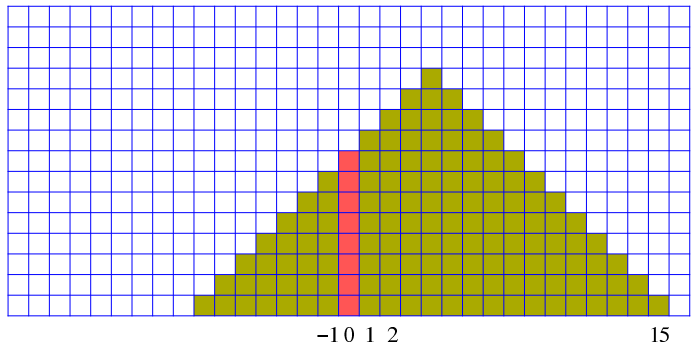}
  \caption{The right-furthest fixed point with $n=144$}
  \label{fig:image1}
\end{figure}

During the $\PSSPM$ evolution we constructed, the original column is always a highest column, and it never receives any grain from its neighborhoods. It would be interesting to investigate the problem in which the positions of the fixed points are considered. In this problem, the fixed points of $\PSSPM$ are not the same as those of $\SSPM$. All the fixed points of $\PSSPM$ might be fully characterized by the furthest fixed points (the maximum and minimum fixed points with respect to the lexicographic order). A possible way to obtain the right-furthest fixed point is that at a current configuration, each column always collapses on the right if it is possible. By doing the experiments on computer, it is surprising that when $n=(8 k+4)^2$ for some $k \in \mathbb{N}$, the furthest fixed point has a nice pyramid-shape which has no plateau and the right-most grain is at distance $11 k+4$ . For example, let $n=144$.  Then the furthest fixed point is illustrated Figure \ref{fig:image1}. It is reasonable to come up with the following conjecture 
\begin{conjec}
$\frac{d(n)}{\sqrt{n}} \sim \frac{11}{8}$, where $d(n)$ denotes the distance of the right-most grain in the right-furthest fixed points to the original column.
\end{conjec}
Up to now, we do not know how to prove the above conjecture or disprove it. Maybe, it needs a deeper analysis on the whole space of reachable configurations of $\PSSPM$.

\bibliographystyle{plain}

\end{document}